\documentclass[reprint,showkeys,showpacs,
superscriptaddress,
 amsmath,
 amssymb,
 aps,
]{revtex4-2}

\usepackage[utf8]{inputenc}
\usepackage{graphicx}% Include figure files
\usepackage{dcolumn}% Align table columns on decimal point
\usepackage{bm}% bold math
\usepackage[colorlinks]{hyperref}
\hypersetup{
    linkcolor = blue,      % 超链接文本颜色
    citecolor = blue,      % 引用标记颜色
    urlcolor  = blue       % URL链接颜色
}
\usepackage{color}
\usepackage{amsmath}

\begin{document}

\title{Interference-induced suppression of particle emission from a Bose-Einstein condensate in lattice with time-periodic modulations}

\author{L. Q. Lai}
\email{lqlai@njupt.edu.cn}
\affiliation{School of Science, Nanjing University of Posts and Telecommunications, Nanjing 210023, China}

\author{Z. Li}
\affiliation{School of Electronic Engineering, Chengdu Technological University, Chengdu 611730, China}

\date{\today}

\begin{abstract}

Emission of matter-wave jets from a parametrically driven condensate has attracted significant experimental and theoretical attention due to the appealing visual effects and potential metrological applications. In this work, we investigate the collective particle emission from a Bose-Einstein condensate confined in a one-dimensional lattice with periodically modulated interparticle interactions. We give the regimes for discrete modes, and find that the emission can be distinctly suppressed. The configuration induces a broad band, but few particles are ejected due to the interference of the matter waves. We further qualitatively model the emission process, and demonstrate the short-time behaviors. This engineering provides a way for manipulating the propagation of particles and the corresponding dynamics of condensates in lattices, and may find use in the dynamical excitation control of other nonequilibrium problems with time-periodic driving.

\end{abstract}



\maketitle

\section{Introduction}

The ability of precisely manipulating the interparticle interactions in many-body systems has provided opportunities to investigate the exotic behaviors of quantum matters and control the quantum engineering \cite{bloch}, as well as demonstrate a variety of fundamental physical problems, such as quantum phase transition and nonequilibrium dynamics \cite{polkovnikov}. Time-periodic driving is one of the major schemes especially implemented in experiments, which opens possibilities of the explorations on fascinating
quantum phenomena \cite{moon,eckardt} and the remarkable applications in a large class of platforms, ranging from topological systems \cite{kitagawa,wang,lu,molignini} to optical lattices \cite{goldman,schweizer,wintersperger,song}.

Coherent manipulation is particularly promising in the gases of ultracold atoms, where the excitation of atoms from a Bose-Einstein condensate can be well controlled via Feshbach resonances \cite{pollack,chin}. Pair atoms shared the energy from the periodic drive, and were ejected in opposite directions after collisions, leading to the collective emission of matter-wave jets resembling fireworks \cite{clark}. A number of subsequent advances revealing new aspects of Bose fireworks and related systems have been reported \cite{fu,zhang,feng,fu1,meznarsic,kim,chen,wu,chih,lellouch,martone, zhang2,xu,liu,lai1,lai2,lai3,lai4}, where in the theoretical side the experimentally accessible lattice models were introduced as a transparent way of studying the relevant physics. The behavior of the trapped atoms was mimicked by the
hopping of particles from the condensate to the leads and between neighboring sites. In addition to the widely studied pair emission, a novel single-particle emission \cite{lai1}, the resonant enhancement \cite{lai2}, the effects of drive imbalance \cite{lai3}, as well as a distinctly intermittent emission \cite{lai4}  have been demonstrated, respectively.

In cold atom experiments, one can configure unconventional lattice structures and control the couplings to reveal novel nonequilibrium effects based on specific demands. There have been seminal works paying attention to the dynamical behaviors of condensates confined in a double-well potential \cite{gu,xue,cataldo,chang,haldar,liang,lindberg,korshynska} and the reverse of quantum dynamics in weakly interacting Bose condensates with time-dependent interactions \cite{walls,bloch1,zoller,linnemann,zhou0,hu1,chen2,zhou1,zhou2,cheng,zhang3}, where the symmetry plays a significant role in the dynamic process, leading to unique geometric interpretations of the instabilities and feasible implementations of tailoring the collective excitations of many-body quantum systems \cite{zhou1,zhou2}. These inspiring results thus naturally raise an interesting question as to whether there exist analogs in such lattice systems, i.e., the possible suppression of the matter-wave jets, which could be conducive to the precise control of the corresponding dynamics. In Ref.~\cite{lai2} we focused
on a periodically driven condensate in a one-dimensional
lattice to demonstrate the parametric resonance, where
a local deep trap confined two central sites, with each
site solely connected to the left or right lead. Here we introduce a similar geometry, however, with some more sophisticated setups, where the central sites are coupled to both leads. We analyze the dynamics within perturbative analysis and figure out the available broad band for discrete modes. When compared with the previous results, we find the strong suppression of particle emission, and present both analytical and numerical calculations to qualitatively model the short-time process.

The paper is organized as follows. In Sec.~\ref{model} we illustrate our theoretical model and the corresponding equations of motion, and use perturbative analysis to discuss the regimes of the discrete modes. In Sec.~\ref{parametric} we compare the solutions from the present work with a previous one to demonstrate the distinct suppression of the particle emission, and further qualitatively clarify the results. A summary is given in Sec.~\ref{summary}.

\section{Model and Perturbation}\label{model}

\begin{figure}[htbp]
\includegraphics[width=\columnwidth]{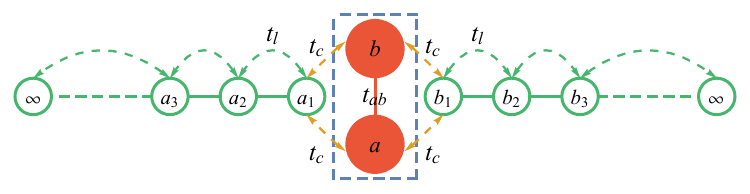}
\caption{(color online) Illustration of the  one-dimensional geometry. The red solid circles symbolize the central sites in a local deep trap denoted by the blue dashed box, which clarifies the couplings of $a$ and $b$ with both $a_{1}$ and $b_{1}$. The green empty circles labeled by $a_{1}(b_{1}),a_{2}(b_{2}),\ldots,\infty$ represent the sites on each lead.}
\label{latticepic}
\end{figure}

The system under consideration is a one-dimensional infinite lattice, as depicted in Fig.~\ref{latticepic}, where a local trapping potential of depth $V$ is placed to confine a Bose-Einstein condensate. The central sites labeled $a$ and $b$ are coupled with amplitude $t_{ab}$, which enables the tunnelling of particles from one site to the other. Atoms with sufficient energy can run away from the trap, while hopping onto either site $a_{1}$ or $b_{1}$ with strength $t_{c}$, and the couplings between neighboring sites in each lead are quantified by $t_{l}$.
The symmetric geometry can be literally realized in an experiment by putting a condensate in a double-well potential combined with optical lattices, while the couplings can be specifically manipulated and the interparticle interactions are modulated by time-periodic driving fields. \cite{lignier,kuhr}

For simplicity, we only take the interactions between
atoms which sit on the central sites into account, and neglect those outside of the trap. Thus the Hamiltonian of such a system reads
\begin{eqnarray}
\hat{H} &=&V\left( \hat{a}_{0}^{\dag }\hat{a}_{0}+\hat{b}_{0}^{\dag }\hat{b}_{0}\right)-t_{ab}\left(\hat{a}_{0}^{\dag }\hat{b}_{0}+\hat{b}_{0}^{\dag }\hat{a}_{0}\right) \nonumber \\
&&+\frac{1}{2}\left[ U+g\left(
t\right) \right] \left(
\hat{a}_{0}^{\dag }\hat{a}_{0}^{\dag }\hat{a}_{0}\hat{a}_{0}
+\hat{b}_{0}^{\dag }\hat{b}_{0}^{\dag }\hat{b}_{0}\hat{b}_{0}
\right) \nonumber \\
&&-t_c \left(\hat{a}_1^\dag \hat{a}_0+\hat{a}_1^\dag \hat{b}_0+\hat{b}_1^\dag \hat{a}_0+\hat{b}_1^\dag \hat{b}_0+\rm{H.c}. \right) \nonumber\\
&&-t_{l}\sum_{j=1}^{\infty }\left( \hat{a}_{j+1}^{\dag }\hat{a}_{j}+\hat{b}_{j+1}^{\dag }\hat{b}_{j}+\rm{H.c.}\right),
\end{eqnarray}
where $\hat{a}_{j}^{\dag}$ and $\hat{b}_{j}^{\dag}$ are the creation bosonic operators on the $j$th site, while $\hat{a}_{0}^{\dag}$ and $\hat{b}_{0}^{\dag}$ correspond to the central sites. A time-independent on-site interaction $U$ and a time-dependent periodic driving $g(t)$ characterize the pairwise particle interactions, for which we take the sinusoidal modulation $g(t)=g\sin(\omega t)$ with $g$ representing the drive strength and $\omega$ being the drive frequency.

The lattice involves the nature of a macroscopic number of particles, and hence we can use a mean-field approximation to replace the operators $\hat{a}$ and $\hat{b}$ with some complex numbers $a$ and $b$, i.e., the absolute squares of the expectation values $a_{j}=\langle \hat{a}_{j}\rangle$ and $b_{j}=\langle \hat{b}_{j}\rangle$ represent the population of atoms on site $j$ in each lead.

Since the dc component of the scattering length was generally kept small in experiments \cite{clark,fu,zhang,feng,fu1} and a finite $U$ did not qualitatively change the relevant physics \cite{lai1}, by neglecting the on-site interaction term and taking units where $\hbar=1$, and according to the equations of motion
$\partial_{t}\hat{A}=i[\hat{H},\hat{A}]$ in the Heisenberg picture,
we thus write down, respectively, for the central sites
\begin{eqnarray}
i\partial_{t}
\left(
\begin{array}{cc}
a_{0} \\ b_{0}
\end{array}
\right)
&=&\left(
\begin{array}{cc}
\Lambda_{a} & 0 \\
0 & \Lambda_{b}
\end{array}
\right)
\left(
\begin{array}{cc}
a_{0} \\ b_{0}
\end{array}
\right)-t_{ab}
\left(
\begin{array}{cc}
b_{0} \\ a_{0}
\end{array}
\right)
-t_{c}\Omega_{1}
\end{eqnarray}
with $\Lambda_{a}=V+g(t)|a_{0}|^2$, $\Lambda_{b}=V+g(t)|b_{0}|^2$ and $\Omega_{m}=
\big(\begin{smallmatrix}
a_{m}+b_{m} \\ a_{m}+b_{m}
\end{smallmatrix}\big)$, and for the sites where $j \geq 1$,
\begin{eqnarray}
i\partial_{t}
\left(
\begin{array}{cc}
a_{1} \\ b_{1}
\end{array}
\right)
&=&-t_{c}\Omega_{0}
-t_{l}
\left(
\begin{array}{cc}
a_{2} \\ b_{2}
\end{array}
\right),  \\
i\partial_{t}
\left(
\begin{array}{cc}
a_{j} \\ b_{j}
\end{array}
\right)
&=&-t_{l}
\left(
\begin{array}{cc}
a_{j+1} \\ b_{j+1}
\end{array}
\right)
-t_{l}
\left(
\begin{array}{cc}
a_{j-1} \\ b_{j-1}
\end{array}
\right).  \label{emoj}
\end{eqnarray}
At time $t<0$, the system is in its equilibrium when the drive is absent. We find the stationary solutions by making the ansatz $a_{0}(t)=\alpha e^{-i \nu t}$, $b_{0}(t)=\beta e^{-i \nu t}$ with $\alpha$ and $\beta$ being constant, and assuming that
\begin{eqnarray}
a_1(t) = \frac{\alpha+\beta}{2} e^{-i \nu t} e^{-\lambda_1}, \,
a_2(t) = \frac{\alpha+\beta}{2} e^{-i \nu t} e^{-\lambda_1-\lambda}.
\end{eqnarray}
The linear Eq.~(\ref{emoj}) leads to $a_j/a_{j-1}=b_{j}/b_{j-1}=e^{-\lambda}$, thus $\cosh\lambda=\nu/(-2 t_l)$, and for $j\geq 1$ we have explicitly
\begin{eqnarray}
 a_{j}(t)=b_{j}(t)=\frac{\alpha+\beta}{2} e^{-i \nu t} e^{-\lambda_1} e^{-\lambda (j-1)}
\end{eqnarray}
with $\lambda_{1}=-{\rm ln}\left[-2t_{c}/(\nu+t_{l} e^{-\lambda})\right]$. Eliminating the leads based on the Green's function technique \cite{lai2}
\begin{eqnarray} \label{aj}
a_{j}(t) &=& b_{j}(t) = -t_{c}\int^{t}d\tau G_{j1}(t-\tau)\left[a_{0}(\tau)+b_{0}(\tau)\right],
\end{eqnarray}
where $G_{j1}(t)=i^{j-2}[jJ_{j}(2t_{l}t)]\theta(t)/(t_{l}t)$ is the time-domain Green's function
% \begin{eqnarray}
% G_{j1}(t)=i^{j-2}\frac{jJ_{j}(2t_{l}t)}{t_{l}t}\theta(t)
% \end{eqnarray}
with $J_{n}(x)$ the Bessel function of the first kind and $\theta(t)$ the step function, we obtain the nonlinear integro-differential equation
\begin{eqnarray}
i\partial_{t}
\left(
\begin{array}{cc}
a_{0} \\ b_{0}
\end{array}
\right)
&=&
\left(
\begin{array}{cc}
\Lambda_{a} & 0 \\
0 & \Lambda_{b}
\end{array}
\right)
\left(
\begin{array}{cc}
a_{0} \\ b_{0}
\end{array}
\right)-t_{ab}
\left(
\begin{array}{cc}
b_{0} \\ a_{0}
\end{array}
\right)\nonumber \\
&&+2t_c^2 \int^t G_{11}(t-\tau)d\tau\Omega_{0}(\tau). \label{integrodifferential}
\end{eqnarray}

% \begin{eqnarray}
% i\partial_t a_{0}&=&Va_{0}+g(t)\vert a_{0} \vert^2 a_{0}-t_{ab} b_{0} \nonumber \\
% &&+2t_c^2 \int^t G_{11}(t-\tau)\left[a_{0}(\tau)+b_{0}(\tau) \right]d\tau , \\
% i\partial_t b_{0}&=&Vb_{0}+g(t)\vert b_{0} \vert^2 b_{0}-t_{ab} a_{0} \nonumber \\
% &&+2t_c^2 \int^t G_{11}(t-\tau)\left[a_{0}(\tau)+b_{0}(\tau) \right]d\tau,
% \end{eqnarray}

To largely simplify the analysis, we treat $g$ as a small parameter, where one could begin by finding the perturbative solution with $g=0$. Accordingly, Eq.~(\ref{integrodifferential}) becomes a linear equation, whose solutions are linear combinations of functions of $a_0$ and $b_0$, with
\begin{eqnarray}
\nu\alpha &=& V\alpha-t_{ab}\beta+2t_c^2G_{11}(\nu)(\alpha+\beta),\\
\nu\beta &=& V\beta-t_{ab}\alpha+2t_c^2G_{11}(\nu)(\alpha+\beta),
\end{eqnarray}
where, as in Ref.~\cite{lai1}, the frequency-domain Green's function reads
\begin{eqnarray}
G_{11}(\epsilon)=\frac{\epsilon}{2t_{l}^2}-i\sqrt{\frac{1}{t_{l}^2}-\frac{\epsilon^2}{4t_{l}^4}}.
\end{eqnarray}

We want the symmetric mode with $\beta=\alpha$ to be stable, which means that there should be a real-value solution to the equation
\begin{eqnarray}
\nu-V+t_{ab}-4t_{c}^2 G_{11}(\nu)=0, \label{eqsym}
\end{eqnarray}
and it specifically constrains the allowed values of the trapping potential $V$. Though the above equation can be readily solved by substituting in the expression of $G_{11}$ and shuffling terms around, it suffices for us to perturbatively solve it. To the zeroth order in $t_{c}$, we obtain an explicit expression
\begin{eqnarray}
\nu_{s}^{(0)}=V-t_{ab},
\end{eqnarray}
while at the second order one simply substitutes this into Eq.~(\ref{eqsym}) to get
\begin{eqnarray}
\nu_{s}^{(2)} &=& V-t_{ab} \nonumber\\
&&+\frac{2t_{c}^2}{t_{l}^2} \left(V-t_{ab}-i \sqrt{4t_{l}^2-(V-t_{ab})^2}\right),
\end{eqnarray}
which leads to the constraints of $|V-t_{ab}|>2 t_{l}$. Specializing to the case where $V=-|V|<0$, in the symmetric mode we subsequently  reach the allowed values of the trapping potential as $|V|+t_{ab}>2t_{l}$.

One can also, in this slick way, obtain the frequency of the antisymmetric mode $\beta=-\alpha$ for both the zeroth and the second order of $t_{c}$, with
\begin{eqnarray}
\nu_{a}^{(0)}=\nu_{a}^{(2)}=V+t_{ab}.
\end{eqnarray}
To observe significant particle emission, we need to parametrically excite the antisymmetric mode, which requires the complex solutions. However, it is obvious that under this circumstance they have explicitly only real values in both orders. As a consequence, we are probably incapable of being in the regime where the symmetric mode is stable but the antisymmetric mode is damped, i.e., the collective excitations of particles could be greatly suppressed, and there exists merely a very weak jet.

\section{Parametric Drive} \label{parametric}

For the verification of the scenario, we now parametrically drive the system and explore its nonlinear dynamics by numerically solving Eq.~(\ref{integrodifferential}). We assume that the weak drive is turned on at time $t=0$. We introduce, without any loss of generality, initially slight difference for $a_{0}(t=0)=\alpha=1.01$ and $b_{0}(t=0)=\beta=1$ to seed the system at the lowest symmetric mode of $\nu_{s}=V-t_{ab}$. We also fix $t_{ab}=1$ as the energy scale of the system in most of the numerics, such that the times are measured in units of $\hbar/t_{ab}$.

\begin{figure}[htbp]
\includegraphics[width=0.9\columnwidth]{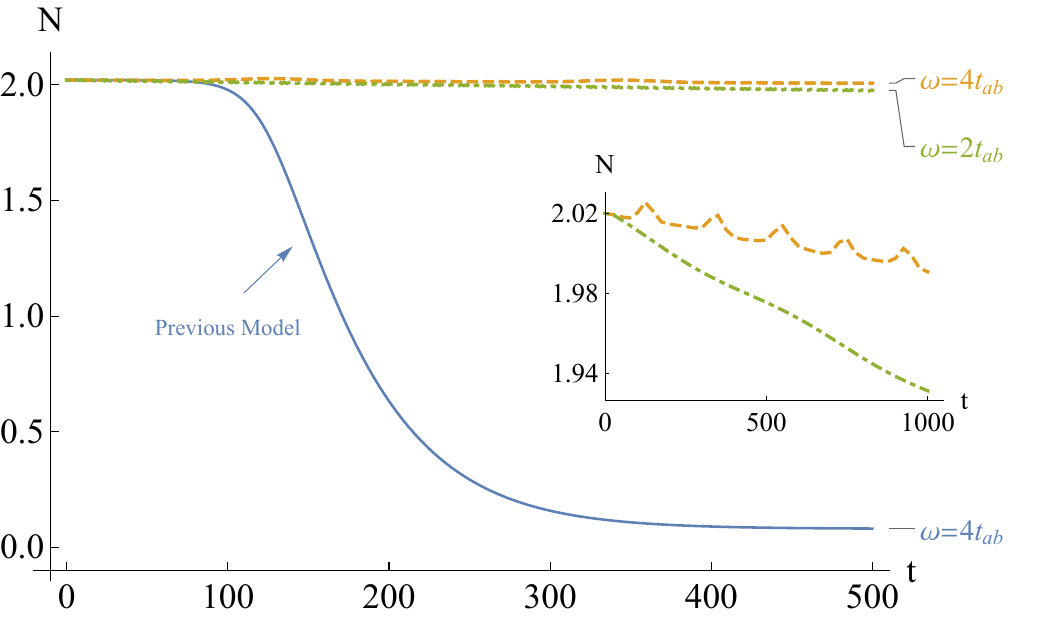}
\caption{(color online) Time dependence of the number of trapped atoms $N$ for both models with typical drive frequencies $\omega$. The blue solid line corresponds to the previous two-site model from Ref.~\cite{lai2} with $\omega=4t_{ab}$, while the green dot-dashed line and the orange dashed line are for the present model with $\omega=2t_{ab}$ and $\omega=4t_{ab}$, respectively. Inset shows a more detailed behavior of the present model with a longer time interval. Here, we have kept the drive strength as $g=0.1$ and the trapping potential as $V=-2$. The coupling strengths are $t_{c}=0.1$ and $t_{l}=1$.}
\label{comparison}
\end{figure}

We first compare the collective emission of particles in the present model with the previous one from Ref.~\cite{lai2}, by showing their decaying behaviors under typical drive frequencies. Specifically they are both two-site infinite lattices with somewhat similar configurations, but the previous one exhibited a resonant enhancement. As can be plainly seen in Fig.~\ref{comparison}, when the drive strength is as small as $g=0.1$ and the drive frequency is $\omega=4t_{ab}$, for the previous model the number of trapped atoms $ N=|a_{0}|^2+|b_{0}|^2$ bears a short plateau around $t<100$ with only a few particles escaped. At intermediate times $150<t<250$, the condensate suddenly starts to eject a burst of atoms until that few are remained. As for the present system, it undergoes a slow but very steady decay under both frequencies of $\omega=2t_{ab}$ and $\omega=4t_{ab}$. The detailed behaviors are characterized in the inset, where a fairly small portion of particles are ejected even in a much longer time interval.

\begin{figure}[htbp]
\includegraphics[width=0.9\columnwidth]{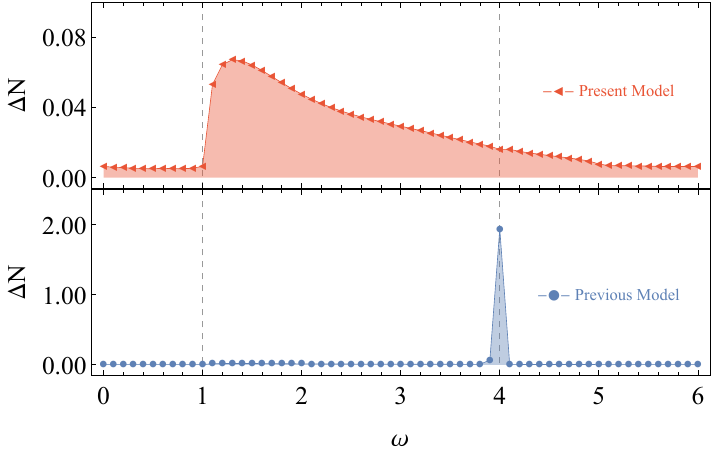}
\caption{(color online) Number of excited atoms $\Delta N$ under different frequency $\omega$, which is calculated from $\Delta N=\left(|a_{0}(t=0)|^2+|b_{0}(t=0)|^2\right)-\left(|a_{0}(t=\tau_{e})|^2+|b_{0}(t=\tau_{e})|^2\right)$, up to time $\tau_{e}=500$. The upper panel corresponds to the present model, while the lower panel comes from the previous model in Ref.~\cite{lai2}. The trapping potential is $V=-2$, and the drive strength is $g=0.1$. The coupling strengths are $t_{c}=0.1$ and $t_{l}=1$.}
\label{frequencysweep}
\end{figure}

\begin{figure}[htbp]
\includegraphics[width=0.9\columnwidth]{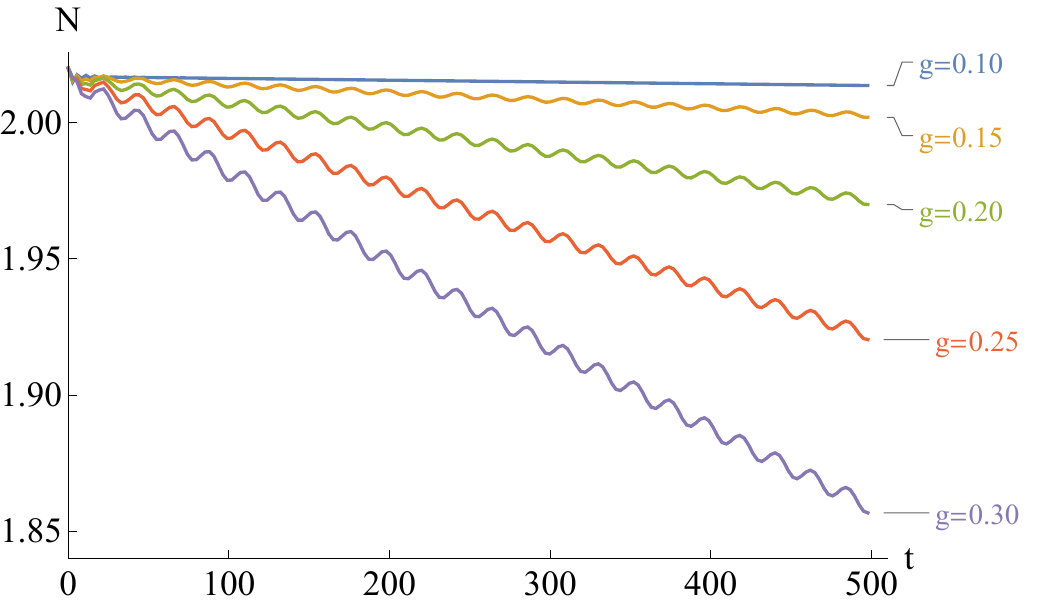}
\caption{(color online) Time dependence of the  number of trapped atoms $N$ for the present model with different drive strength $g$. Here, the drive frequency is fixed as $\omega=2t_{ab}$, and we have taken $V=-2$, $t_{c}=0.1$ and $t_{l}=1$.}
\label{drivecomparison}
\end{figure}

To check that whether the discrepancies shown above are induced by the selection of certain drive frequencies, in Fig.~\ref{frequencysweep} we keep the drive strength as $g=0.1$ while sweeping the frequency, and also compare the corresponding results from Ref.~\cite{lai2}. For the previous model, a distinct peak appears at $\omega= 4t_{ab}$ with a typical narrow bandwidth, while for other non-resonant frequencies there are hardly ejected atoms. In sharp contrast, a much broader band emerges roughly from $\omega=t_{ab}$ to $\omega=4t_{ab}$ for the present model, however, with very few excited atoms. Even when we tune to a relatively larger drive strength (e.g. $g=0.3$), as shown in Fig.~\ref{drivecomparison}, the number $N$ can oscillate and fall with a greater extent, but there seems to be still a small particle jet rather than large pulse, which indicates that with such a configuration the collective emission is distinctly suppressed.

For the two-site lattice in the previous model, once the time-periodic modulation with an appropriate frequency that is within the narrow resonance region is turned on, the condensate wave functions undergo rapid oscillations and the atoms in the ground state are subsequently pumped to a higher energy level corresponding to the antisymmetric mode. The excited atoms with sufficient energy escape from the trap and travel along the leads, exhibiting a parametric resonance. With respect to the present configuration, however, the main difference from the previous model lies in Eq.~(\ref{aj}), where two branches of the matter waves coming from the central sites $a$ and $b$ tunnel and interfere due to the sophisticated hoppings. One could further refer to the coherent destruction of tunneling for a more insightful perspective, where the suppression of tunneling in the periodically driven two-level system may be viewed as the destructive interference between transition paths. \cite{grossmann1,grossmann2,kayanuma,della,creffield} The decay rate of the particles in the source condensate is thus far inferior to the growth rate, such that the antisymmetric mode can hardly leak into the leads, and there are no longer significant jets.

To demonstrate the formalism, in the limit where $t_{c}$ is negligible we qualitatively model the emission process with the help of the method of multiple scales, where
\begin{eqnarray}
a_{0}(t)&=&e^{-i \nu_{s} t} u(t) + e^{-i \nu_{a} t} v(t), \label{ansatza} \\
b_{0}(t)&=&e^{-i \nu_{s} t} u(t) - e^{-i \nu_{a} t} v(t), \label{ansatzb}
\end{eqnarray}
with $u(t)$ and $v(t)$ both slowly varying, and $\nu_{(s,a)}$ denoting the zeroth-order frequencies corresponding to the discrete modes. Substituting the ansatz into Eq.~(\ref{integrodifferential}) and collecting the terms leads to (see Appendix~\ref{mms})
\begin{eqnarray}
\partial_{t} u &=& 2gX(t)v^{2}u, \label{ana1} \\
\partial_{t} v &=& -2gX(t)u^{2}v, \label{ana2}
\end{eqnarray}
where $X(t)=\sin(\omega t)\sin(4t_{ab}t)$. Since $u^2$ and $v^2$ represents the number of particles in the symmetric mode and in the antisymmetric mode, respectively, if one linearizes about $v=0$, i.e., all particles are initially placed in the ground state, from the above equations we find $v$ decays to zero while $u$ grows. In other words, we are not able to produce large particle jets by leaking the antisymmetric mode into the lead.

\begin{figure}[htbp]
\includegraphics[width=\columnwidth]{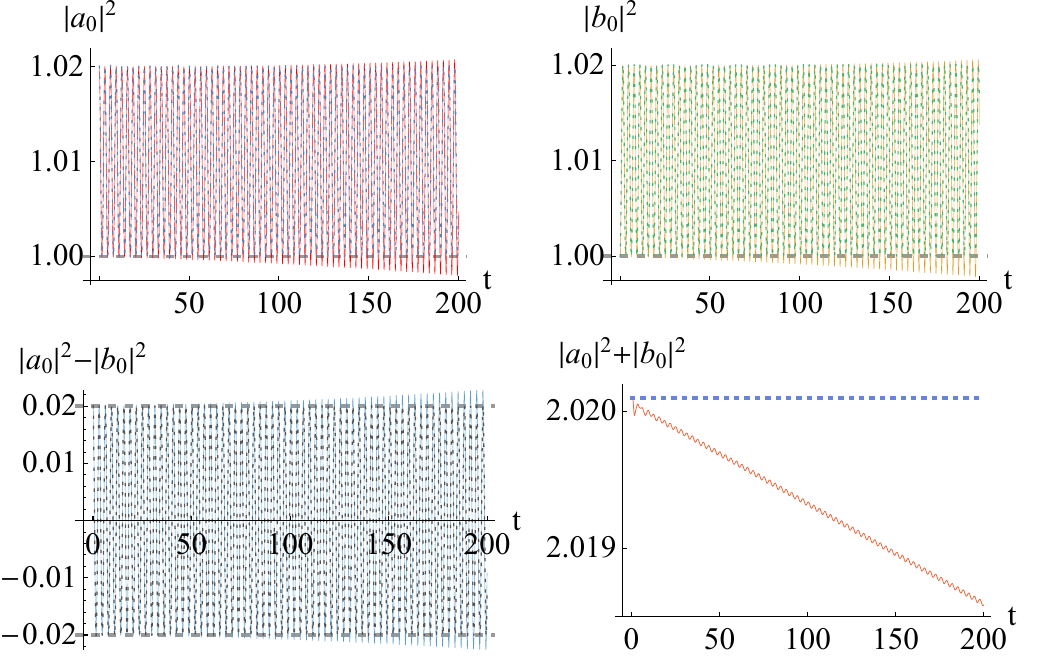}
\caption{(color online) Typical examples of the time evolution of the particle number $|a_{0}|^2$ and $|b_{0}|^2$ on both central sites, the imbalance $|a_{0}|^2-|b_{0}|^2$ and the total particle number in the condensate $|a_{0}|^2+|b_{0}|^2$. The analytical solutions (dotted line) and the numerical results (solid line) are nearly indistinguishable at short times. Here, the drive strength is $g=0.1$, the drive frequency is fixed as $\omega=2t_{ab}$ and the coupling strength is tuned to $t_{c}=0.01$. We have also taken $V=-2$ and $t_{l}=1$.}
\label{comparetot}
\end{figure}

Figure~\ref{comparetot} compares the analytical solutions coming from Eqs.~(\ref{ana1}) and (\ref{ana2}) with the numerical result. Both outcomes clearly describe the rapid oscillation and slow time evolution of the short-time behaviors of the system, and they fit quite well at rough times $t<80$. For longer intervals, the deviations emerge, where the numerics show that the particle imbalance $|a_{0}|^2-|b_{0}|^2$ oscillates around and grows up with time rather than falls, under which the system maintains in the stage of build-up, corresponding to the fact that the total particle number $|a_{0}|^2+|b_{0}|^2$ from the numerics drops very slightly.

\begin{figure}[htbp]
\includegraphics[width=\columnwidth]{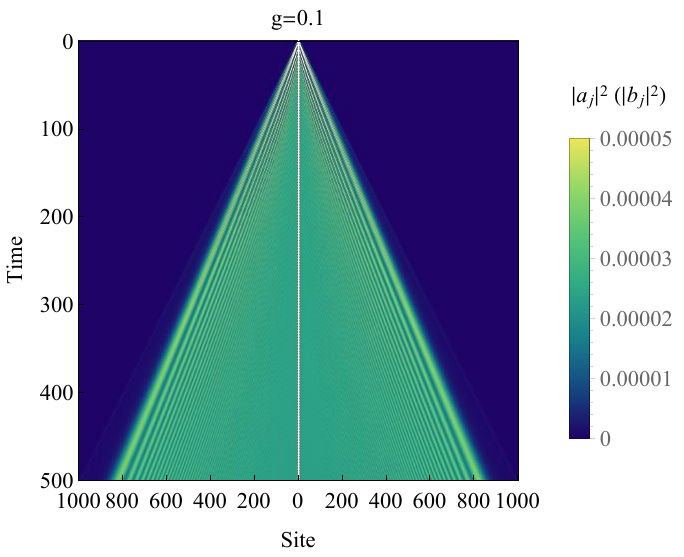}
\caption{(color online) Number of the particles on the $j$th site of each lead as a function of time. Here, we have taken $V=-2$, $g=0.1$, $\omega=2t_{ab}$, $t_{c}=0.1$ and $t_{l}=1$. Note the scaled legend.}
\label{jetarray}
\end{figure}

There can always be certain oscillations and ejected particles from the condensate once the time-periodic driving field is applied, while the analytical formalsim, which is limited in the regime where the drive strength $g$ and the coupling strength $t_{c}$ are both negligible, demonstrates the approximate stability throughout. Therefore, for a larger drive the linearization may be no longer justified. We further illustrate the detailed structure of a jet by plotting the number of particles on each site of the leads, as shown in Fig.~\ref{jetarray}. Note that the jets can be moderately visible and continuous here, however, with fairly few ejected particles, which bears significant difference from the previous pulse.

\section{Summary}\label{summary}
We have introduced a one-dimensional lattice, where the trapped central sites are coupled with both leads, to investigate the collective emission of particles from a Bose-Einstein condensate with time-periodic modulations. Perturbative analysis is used for simplification, for which we validate through numerical calculations.

In a previous work where the configuration was somewhat similar, the collective modes resulted in a dramatic enhancement such that a small drive could induce large pulses when the drive frequency was tuned to resonance \cite{lai2}. In the present model, however, a much broader band emerges with only few particles ejected even when the drive strengths are increased, i.e., the emission is greatly suppressed. This phenomenon might be attributed to the destructive interference of the outcoming matter waves from the central sites, leading to the decay rate of the particles being much lower than the growth rate, and there are insignificant particle jets outside of the trap.

The symmetric geometry and the couplings can be readily implemented with optical lattices in experiments. One could, based on particular purposes, utilize specific engineering to either enhance or suppress the dynamic processes, and such framework would enable an intensive study of the correlations among the particle emission, the couplings and the various leads.  The ability of tailoring the dynamical excitation and coherently steering particles to produce directed motion in a lattice system also points to the possible implementation in coupled quantum dots and quantum information processing.

\section*{Acknowledgements}
L.Q.L. would like to thank Professor Erich J. Mueller for the profound instructions. This work was supported by the China Scholarship Council (Grant No.~201906130092), the Natural Science Research Start-up Foundation of Recruiting Talents of Nanjing University of Posts and Telecommunications (Grant No.~NY223065), and the Natural Science Foundation of Sichuan Province (Grant No.~2023NSFSC1330).

\begin{appendix}
\section{Derivation of the analytical relation}\label{mms}
Here we utilize the method of multiple scales to reproduce the results from Sec.~\ref{parametric}, in the limit where the drive strength $g$ and the coupling strength $t_{c}$ are small. Substituting Eqs.~(\ref{ansatza}) and (\ref{ansatzb}) into Eq.~(\ref{integrodifferential}), and neglecting the slow variation of any function $f(t)$ in the integral by
\begin{eqnarray}
\int^t G_{11}(t-\tau)f(\tau)e^{-i \xi \tau} d\tau
\approx f(t)e^{-i \xi t} G_{11}(\xi),
\end{eqnarray}
as well as using the neat combinations
\begin{eqnarray}
&&g \sin(\omega t) (|a_{0}|^2 a_{0}+|b_{0}|^2 b_{0})
=\frac{g}{i} \left(e^{i \omega t}-e^{-i \omega t}\right)\times \nonumber \\
&&\left(
e^{-i \nu_s t} ( |u|^2+2 |v|^2)  u
+ e^{i (\nu_s-2 \nu_a) t} v^2 u^*
\right),\\
&& g \sin(\omega t) (|a_{0}|^2 a_{0}-|b_{0}|^2 b_{0})
=\frac{g}{i} \left(e^{i \omega t}-e^{-i \omega t}\right)\times \nonumber \\
&&\left(
e^{-i \nu_a t} (2 |u|^2+ |v|^2) v
+ e^{i (\nu_a-2 \nu_s) t} u^2 v^*
\right),
\end{eqnarray}
we thus write down the equations of motion for $u$ and $v$,
\begin{eqnarray}
i \partial_t \frac{a_{0}+b_{0}}{2} &=& (\nu_{s}u+i\partial_{t}u) e^{-i\nu_{s}t}  \\
&=& (V-t_{ab})u e^{-i\nu_{s}t}+
\frac{g\sin(\omega t)}{2} (|a_{0}|^2 a_{0}+|b_{0}|^2 b_{0}) \nonumber \\
&&+2t_c^2 \int^{t}G_{11}(t-\tau)e^{-i\nu_{s}\tau}u(\tau), \nonumber \\
i \partial_{t} \frac{a_{0}-b_{0}}{2} &=& (\nu_{a}v+i\partial_{t}v) e^{-i\nu_{a}t}  \\
&=&(V+t_{ab})v e^{-i\nu_{a}t}+
\frac{g\sin(\omega t)}{2} (|a_{0}|^2 a_{0}-|b_{0}|^2 b_{0}), \nonumber
\end{eqnarray}
which can be transformed into
\begin{eqnarray}
\partial_{t}u &=& -2it_{c}^2 G_{11}(\nu_{s})u-\frac{g}{2}\left[e^{i\omega t}(|u|^2+2|v|^2)u \right. \nonumber \\
&& \left.+e^{i(\omega+2\nu_{s}-2\nu_{a})t}v^2 u^*
-e^{-i\omega t}(2|u|^2+|v|^2)v \right. \nonumber \\
&&\left.-e^{i(\omega+2\nu_{a}-2\nu_{s})}u^2v^*\right], \label{partialu} \\
\partial_{t}v &=& -\frac{g}{2}\left[e^{i\omega t}(2|u|^2+|v|^2)v+e^{i(\omega+2\nu_{a}-2\nu_{s})t}u^2v^* \right. \nonumber \\
&& \left.-e^{-i\omega t}(2|u|^2+|v|^2)v-e^{-i(\omega+2\nu_{s}-2\nu_{a})t}u^2v^*\right]. \label{partialv}
\end{eqnarray}
Through $u^{*}\times \mathrm{Eq.~(\ref{partialu})} +u \times \mathrm{Eq.~(\ref{partialu})}^{*}$ and $v^{*}\times \mathrm{Eq.~(\ref{partialv})} +v \times \mathrm{Eq.~(\ref{partialv})}^{*}$, we further reach
\begin{eqnarray}
\partial_{t}|u|^2 &=& 4t_{c}^2{\rm{Im}}G_{11}(\nu_{s}) |u|^2-\frac{g}{2}\left[e^{i(\omega+2\nu_{s}-2\nu_{a})t}v^2u^{*2} \right. \nonumber \\
&& \left.+e^{-i(\omega+2\nu_{s}-2\nu_{a})t}v^2|u|^2-e^{i(\omega+2\nu_{a}-2\nu_{s})t}v^2|u|^2 \right. \nonumber \\
&& \left. -e^{-i(\omega+2\nu_{a}-2\nu_{s})t}v^2u^{*2}\right], \\
\partial_{t}|v|^2 &=& -\frac{g}{2}\left[e^{i(\omega+2\nu_{a}-2\nu_{s})t}u^2v^{*2}+e^{-i(\omega+2\nu_{a}-2\nu_{s})t}u^{*2}v^2 \right. \nonumber \\
&&\left.-e^{i(\omega+2\nu_{s}-2\nu_{a})t}u^{*2}v^2-e^{-i(\omega+2\nu_{s}-2\nu_{a})t}u^2v^{*2}\right]. \nonumber \\
\end{eqnarray}
A nice simplifying assumption is to treat $u$ and $v$ as real, we find
\begin{eqnarray}
\partial_{t}u &=& 2g X(t) v^{2}u, \\
\partial_{t}v &=& -2g X(t) u^{2}v,
\end{eqnarray}
where $X(t)=\sin(\omega t) \sin(4t_{ab}t)$. The above equations can readily be numerically solved. Conversely, one could also rewrite from Eqs.~(\ref{ansatza}) and (\ref{ansatzb}) that
\begin{eqnarray}
u e^{-i \nu_{s} t}&=&
\frac{a_{0}+b_{0}}{2},  \\
v e^{-i \nu_{a} t}&=&\frac{a_{0}-b_{0}}{2},
\end{eqnarray}
which can be reconstructed to obtain the particle imbalance
\begin{eqnarray}
|a_{0}|^2-|b_{0}|^2 &=& 2\left[uv^* e^{-i(\nu_{a}-\nu_{s})t}+u^* v e^{i (\nu_{a}-\nu_{s})t}\right] \nonumber \\
&=& 4 u v \cos(2t_{ab}t),
\end{eqnarray}
and the number of total particles
\begin{eqnarray}
|a_{0}|^2+|b_{0}|^2 &=& 2 (|u|^2+|v|^2).
\end{eqnarray}

\end{appendix}


\begin{thebibliography}{99}





\bibitem{bloch} Bloch I, Dalibard J and Zwerger W \href{https://journals.aps.org/rmp/abstract/10.1103/RevModPhys.80.885}{2008 {\it{Rev. Mod. Phys.}} {\bf 80} 885}

\bibitem{polkovnikov} Polkovnikov A, Sengupta K, Silva A and Vengalattore M \href{https://journals.aps.org/rmp/abstract/10.1103/RevModPhys.83.863}{ 2011 {\it{Rev. Mod. Phys.}} {\bf 83} 863}

\bibitem{moon} Moon G, Heo M S, Kim Y, Noh H R and Jhe W \href{https://www.sciencedirect.com/science/article/abs/pii/S0370157317301990?via%3Dihub}{ 2017 {\it{Phys. Rep.}} {\bf 698} 1}

\bibitem{eckardt} Eckardt A \href{https://journals.aps.org/rmp/abstract/10.1103/RevModPhys.89.011004}{ 2017 {\it{Rev. Mod. Phys.}} {\bf 89} 011004}

\bibitem{kitagawa} Kitagawa T, Berg E, Rudner M and Demler E \href{https://journals.aps.org/prb/abstract/10.1103/PhysRevB.82.235114}{ 2010 {\it{Phys. Rev. B}} {\bf 82} 235114}

\bibitem{wang} Wang Y H, Steinberg H, Jarillo-Herrero P and Gedik N \href{https://www.science.org/doi/10.1126/science.1239834}{ 2013 {\it{Science}} {\bf 342} 453}

\bibitem{lu} Lu L, Joannopoulos J D and Solja\v{c}i\'{c} M \href{https://www.nature.com/articles/nphoton.2014.248}{ 2014 {\it{Nat. Photon.}} {\bf 8} 821}

\bibitem{molignini} Molignini P, Chitra R and Chen W \href{https://iopscience.iop.org/article/10.1209/0295-5075/128/36001}{ 2019 {\it{Europhys. Lett.}} {\bf 128} 36001}

\bibitem{goldman} Goldman N and Dalibard J \href{https://journals.aps.org/prx/abstract/10.1103/PhysRevX.4.031027}{ 2014 {\it{Phys. Rev. X}} {\bf 4} 031027}

\bibitem{schweizer} Schweizer C, Grusdt F, Berngruber M, Barbiero L, Demler E, Goldman N, Bloch I and Aidelsburger M \href{https://www.nature.com/articles/s41567-019-0649-7}{ 2019 {\it{Nat. Phys.}} {\bf 15} 1168}

\bibitem{wintersperger} Wintersperger K, Bukov M, N\"{a}ger J, Lellouch S, Demler E, Schneider U, Bloch I, Goldman N and Aidelsburger M \href{https://journals.aps.org/prx/abstract/10.1103/PhysRevX.10.011030}{ 2020 {\it{Phys. Rev. X}} {\bf 10} 011030}

\bibitem{song} Song B, Dutta S, Bhave S, Yu J C, Carter E, Cooper N and Schneider U \href{https://www.nature.com/articles/s41567-021-01476-w}{ 2022 {\it{Nat. Phys.}} {\bf 18} 259}

\bibitem{pollack} Pollack S E, Dries D, Hulet R G, Magalh\~{a}es K M F, Henn E A L, Ramos E R F, Caracanhas M A and Bagnato V S \href{https://journals.aps.org/pra/abstract/10.1103/PhysRevA.81.053627}{ 2010 {\it{Phys. Rev. A}} {\bf 81} 053627}

\bibitem{chin} Chin C, Grimm R, Julienne P and Tiesinga E \href{https://journals.aps.org/rmp/abstract/10.1103/RevModPhys.82.1225}{ 2010 {\it{Rev. Mod. Phys.}} {\bf 82} 1225}

\bibitem{clark} Clark L W, Gaj A, Feng L and Chin C \href{https://www.nature.com/articles/nature24272}{ 2017 {\it{Nature}} {\bf 551} 356}

\bibitem{fu} Fu H, Feng L, Anderson B M, Clark L W, Hu J, Andrade J W, Chin C and Levin K \href{https://journals.aps.org/prl/abstract/10.1103/PhysRevLett.121.243001}{ 2018 {\it{Phys. Rev. Lett.}} {\bf 121} 243001}

\bibitem{zhang} Zhang Z, Yao K X, Feng L, Hu J and Chin C \href{https://www.nature.com/articles/s41567-020-0839-3}{ 2020 {\it{Nat. Phys.}} {\bf 16} 652 (2020)}

\bibitem{feng} Feng L, Hu J, Clark L W and Chin C \href{https://www.science.org/doi/10.1126/science.aat5008}{ 2019 {\it{Science}} {\bf 363} 521}

\bibitem{fu1} Fu H, Zhang Z, Yao K X, Feng L, Yoo J, Clark L W, Levin K and Chin C \href{https://journals.aps.org/prl/abstract/10.1103/PhysRevLett.125.183003}{ 2020 {\it{Phys. Rev. Lett.}} {\bf 125} 183003}

\bibitem{meznarsic} Me\v{z}nar\v{s}i\v{c} T, \v{Z}itko R, Arh T, Gosar K, Zupani\v{c} E and Jegli\v{c} P \href{https://journals.aps.org/pra/abstract/10.1103/PhysRevA.101.031601}{ 2020 {\it{Phys. Rev. A}} {\bf 101} 031601} 

\bibitem{kim} Kim K, Hur J, Huh S J, Choi S and Choi J Y \href{https://journals.aps.org/prl/abstract/10.1103/PhysRevLett.127.043401}{ 2021 {\it{Phys. Rev. Lett.}} {\bf 127} 043401}

\bibitem{chen} Chen T and Yan B \href{https://journals.aps.org/pra/abstract/10.1103/PhysRevA.98.063615}{ 2018 {\it{Phys. Rev. A}} {\bf 98} 063615}

\bibitem{wu} Wu Z G and Zhai H \href{https://journals.aps.org/pra/abstract/10.1103/PhysRevA.99.063624}{ 2019 {\it{Phys. Rev. A}} {\bf 99} 063624}

\bibitem{chih} Chih L Y and Holland M \href{https://iopscience.iop.org/article/10.1088/1367-2630/ab7140}{ 2020 {\it{New J. Phys.}} {\bf 22} 033010}

\bibitem{lellouch} Lellouch S and Goldman N \href{https://iopscience.iop.org/article/10.1088/2058-9565/aab2b9}{ 2018 {\it{Quantum Sci. Technol.}} {\bf 3} 024011}

\bibitem{martone} Martone G I, Larr\'{e} P \'{E}, Fabbri A and Pavloff N \href{https://journals.aps.org/pra/abstract/10.1103/PhysRevA.98.063617}{ 2018 {\it{Phys. Rev. A}} {\bf 98} 063617}

\bibitem{zhang2} Zhang P F and Gu Y F \href{https://scipost.org/SciPostPhys.9.5.079}{ 2020 {\it{SciPost Phys.}} {\bf 9} 079}

\bibitem{xu} Xu P and Zhang W X \href{https://journals.aps.org/pra/abstract/10.1103/PhysRevA.104.023324}{ 2021 {\it{Phys. Rev. A}} {\bf 104} 023324}

\bibitem{liu} Liu N and Tu Z C \href{https://iopscience.iop.org/article/10.1088/1572-9494/abb7f0}{ 2020 {\it{Commun. Theor. Phys.}} {\bf 72} 125501}

\bibitem{lai1} Lai L Q, Yu Y B and Mueller E J \href{https://journals.aps.org/pra/abstract/10.1103/PhysRevA.104.033308}{ 2021 {\it{Phys. Rev. A}} {\bf 104} 033308}

\bibitem{lai2} Lai L Q, Yu Y B and Mueller E J \href{https://journals.aps.org/pra/abstract/10.1103/PhysRevA.106.033302}{ 2022 {\it{Phys. Rev. A}} {\bf 106} 033302}

\bibitem{lai3} Lai L Q and Li Z \href{https://iopscience.iop.org/article/10.1088/1674-1056/ad1172}{ 2024 {\it{Chin. Phys. B}} {\bf{33}} 030308}

\bibitem{lai4} Lai L Q, Li Z, Liu Q H and Yu Y B \href{https://onlinelibrary.wiley.com/doi/full/10.1002/andp.202300365}{ 2024 {\it{Ann. Phys. (Berlin)}} {\bf{536}} 2300365}


\bibitem{gu}Zheng H L and Gu Q \href{https://link.springer.com/article/10.1007/s11467-013-0321-0}{ 2013 {\it{Front. Phys.}} {\bf 8} 375}

\bibitem{xue}Xue R, Li W D and Liang Z X \href{https://cpl.iphy.ac.cn/EN/abstract/abstract58639.shtml#}{2014 {\it{Chin. Phys. Lett.} {\bf 31}} 030302}

\bibitem{cataldo}Cataldo H M and Jezek D M \href{https://journals.aps.org/pra/abstract/10.1103/PhysRevA.90.043610}{2014 {\it{Phys. Rev. A}} {\bf 90} 043610}

\bibitem{chang}Chang N N, Yu Z F, Zhang A X and Xue J K \href{https://iopscience.iop.org/article/10.1088/1674-1056/26/11/115202}{2017 {\it{Chin. Phys. B}} {\bf 26} 115202}

\bibitem{haldar}Haldar S K and Alon O E \href{https://iopscience.iop.org/article/10.1088/1367-2630/ab4315}{ 2019 {\it{New J. Phys.}} {\bf 21} 103037}

\bibitem{liang}Liu J L and Liang J Q \href{https://iopscience.iop.org/article/10.1088/1674-1056/ab44b6}{2019 {\it{Chin. Phys. B}} {\bf 28} 110304}

\bibitem{lindberg}Lindberg D R, Gaaloul N, Kaplan L, Williams J R, Schlippert D, Boege P, Rasel E M and Bondar D I \href{https://iopscience.iop.org/article/10.1088/1361-6455/acae50}{2023 {\it J. Phys. B: At. Mol. Opt. Phys.} {\bf 56} 025302}

\bibitem{korshynska}Korshynska K and Ulbricht S \href{https://journals.aps.org/pra/abstract/10.1103/PhysRevA.109.043321}{2024 {\it{Phys. Rev. A}} {\bf 109} 043321}


\bibitem{walls} Milburn G J, Corney J, Wright E M and Walls D F \href{https://journals.aps.org/pra/abstract/10.1103/PhysRevA.55.4318}{ 1997 {\it{Phys. Rev. A}} {\bf{55}} 4318}

\bibitem{bloch1} Greiner M, Mandel O, H?nsch T W and Bloch I \href{https://www.nature.com/articles/nature00968}{ 2002 {\it{Nature}} {\bf{419}} 51}

\bibitem{zoller} Schachenmayer J, Daley A J and Zoller P \href{https://journals.aps.org/pra/abstract/10.1103/PhysRevA.83.043614}{ 2011 Phys. Rev. A {\bf{83}} 043614}

\bibitem{linnemann} Linnemann D, Strobel H, Muessel W, Schulz J, Lewis-Swan R J, Kheruntsyan K V and Oberthaler M K \href{https://journals.aps.org/prl/abstract/10.1103/PhysRevLett.117.013001}{ 2016 {\it{Phys. Rev. Lett.}} {\bf{117}} 013001}

\bibitem{zhou0} Zhou T, Yang K, Zhu Z, Yu X, Yang S, Xiong W, Zhou X, Chen X, Li C, Schmiedmayer J, Yue X and Zhai Y \href{https://journals.aps.org/pra/abstract/10.1103/PhysRevA.99.013602}{ 2019 {\it{Phys. Rev. A}} {\bf{99}} 013602}

\bibitem{hu1} Hu J, Feng L, Zhang Z and Chin C \href{https://www.nature.com/articles/s41567-019-0537-1}{ 2019 {\it{Nat. Phys.}} {\bf{15}} 785}

\bibitem{chen2} Chen Y Y, Zhang P F, Zheng W, Wu Z G and Zhai H \href{https://journals.aps.org/pra/abstract/10.1103/PhysRevA.102.011301}{ 2020 {\it{Phys. Rev. A}} {\bf 102} 011301}

\bibitem{zhou1} Lyu C, Lv C and Zhou Q \href{https://journals.aps.org/prl/abstract/10.1103/PhysRevLett.125.253401}{ 2020 {\it{Phys. Rev. Lett.}}
{\bf{125}} 253401}

\bibitem{zhou2} Lv C, Zhang R and Zhou Q \href{https://journals.aps.org/prl/abstract/10.1103/PhysRevLett.125.253002}{ 2020 {\it{Phys. Rev. Lett.}} {\bf{125}} 253002}

\bibitem{cheng} Cheng Y T and Shi Z Y \href{https://journals.aps.org/pra/abstract/10.1103/PhysRevA.104.023307}{ 2021 {\it{Phys. Rev. A}} {\bf 104} 023307}

\bibitem{zhang3} Zhang J, Yang X, Lv C, Ma S and Zhang R \href{https://journals.aps.org/pra/abstract/10.1103/PhysRevA.106.013314}{ 2022 {\it{Phys. Rev. A}} {\bf{106}} 013314}


\bibitem{lignier} Lignier H, Sias C, Ciampini D, Singh Y, Zenesini A, Morsch O and Arimondo E \href{https://journals.aps.org/prl/abstract/10.1103/PhysRevLett.99.220403}{ 2007 {\it{Phys. Rev. Lett.}} {\bf 99} 220403}

\bibitem{kuhr} Kuhr S \href{https://academic.oup.com/nsr/article/3/2/170/2460374?login=false}{ 2016 {\it{Nat. Sci. Rev.}} {\bf 3} 170}


\bibitem{grossmann1} Grossmann F, Dittrich T, Jung P and H\"{a}nggi P \href{https://journals.aps.org/prl/abstract/10.1103/PhysRevLett.67.516}{1991 {\it{Phys. Rev. Lett.}} {\bf 67} 516}


\bibitem{grossmann2} Grossmann F and H\"{a}nggi P \href{https://iopscience.iop.org/article/10.1209/0295-5075/18/7/001}{1992 {\it{Europhys. Lett.}} {\bf 18} 571}

\bibitem{kayanuma}Kayanuma Y \href{https://journals.aps.org/pra/abstract/10.1103/PhysRevA.50.843}{1994 {\it{Phys. Rev. A}} {\bf 50} 843}

\bibitem{della}Della Valle G, Ornigotti M, Cianci E, Foglietti V, Laporta P and Longhi S \href{https://journals.aps.org/prl/abstract/10.1103/PhysRevLett.98.263601}{2007 {\it{Phys. Rev. Lett.}} {\bf 98} 263601}

\bibitem{creffield}Creffield C E \href{https://journals.aps.org/prl/abstract/10.1103/PhysRevLett.99.110501}{2007 {\it{Phys. Rev. Lett.}} {\bf 99} 110501}






\end{thebibliography}
\end{document}